\documentclass[letterpaper,english,aps,prl,twocolumn,amsfonts,amssymb,amsmath,longbibliography,superscriptaddress]{revtex4-2}
\usepackage[utf8]{inputenc}
\usepackage{graphicx}
\usepackage{bbold}
\usepackage{xcolor}
\usepackage[mathlines]{lineno}

\DeclareMathOperator\Tr{Tr}

\newcommand{\be}{\begin{equation}}
	\newcommand{\ee}{\end{equation}}
\newcommand{\bspl}{\begin{split}}
	\newcommand{\espl}{\end{split}}
\newcommand{\bea}{\begin{eqnarray}}
	\newcommand{\eea}{\end{eqnarray}}

\newcommand{\angstrom}{\mbox{\normalfont\AA}}

\def\bk{{\bf k}}

\def\lb{\label}
\def\pref#1{(\ref{#1})}

\newcount\bozza \bozza=0
\ifnum\bozza=1
\newdimen\shift \shift=-2truecm
\def\lb#1{%
	{\label{#1}\rlap{\kern\shift{$\scriptstyle#1$}}}}
\else\def\lb#1{\label{#1}} \fi

\begin{document}
	
	\title{Investigating Josephson Plasmons in Layered Cuprates via non-linear THz spectroscopy}
	
	\author{Jacopo Fiore}
	\email{jacopo.fiore@uniroma1.it}
	\affiliation{Department of Physics and ISC-CNR, ``Sapienza'' University of Rome, P.le
		A. Moro 5, 00185 Rome, Italy}
	\author{Niccolò Sellati}
	\affiliation{Department of Physics and ISC-CNR, ``Sapienza'' University of Rome, P.le
		A. Moro 5, 00185 Rome, Italy}
	\author{Francesco Gabriele}
	\affiliation{Department of Physics and ISC-CNR, ``Sapienza'' University of Rome, P.le
		A. Moro 5, 00185 Rome, Italy}
	\author{Claudio Castellani}
	\affiliation{Department of Physics and ISC-CNR, ``Sapienza'' University of Rome, P.le
		A. Moro 5, 00185 Rome, Italy}
	\author{Goetz Seibold}
	\affiliation{Institut f\"ur Physik, BTU Cottbus-Senftenberg, PBox 101344, 03013 Cottbus,
		Germany}
	\author{Mattia Udina}
	\affiliation{Department of Physics and ISC-CNR, ``Sapienza'' University of Rome, P.le
		A. Moro 5, 00185 Rome, Italy}
	\author{Lara Benfatto}
	\email{lara.benfatto@roma1.infn.it}
	\affiliation{Department of Physics and ISC-CNR, ``Sapienza'' University of Rome, P.le
		A. Moro 5, 00185 Rome, Italy}
	
	\begin{abstract}
	Josephson plasmons in layered superconductors represent a natural source of optical non-linearity, thanks to their intrinsically anharmonic nature. Here we derive the selection rules behind non-linear plasmonics showing its dependence on plasmonic branches hidden to other spectroscopies, like RIXS. We benchmark our results for the case of layered cuprates, showing how in a layered system the combined effect of plasmon dispersion and light polarization can move the resonance of the bilayer system away from the plasma edge measured in linear spectroscopy. Our results demonstrate the dependence of the non-linear THz response on the convoluted plasmon dispersion in a momentum region complementary to RIXS, and offer a possible perspective for the generation of THz pulses by artificially designed Josephson heterostructures.

	
	\end{abstract}
	\date\today
	\maketitle
	
	{\em Introduction}.	
Plasma modes describe the propagation of longitudinal electromagnetic waves in metals, hybridized with electronic charge fluctuations. As such, they can be detected via density-sensitive probes, like inelastic X-ray scattering\cite{ament_Rev.Mod.Phys.11} (RIXS) or electron energy-loss spectroscopy\cite{garciadeabajo_Rev.Mod.Phys.10} (EELS). Whenever the metal becomes superconducting (SC) the fluctuations of the SC phase, canonically conjugate to the density, bring up the information on the plasma mode\cite{nagaosa_99}.

\begin{figure}[tb]
		\centering
		\includegraphics[width=\columnwidth]{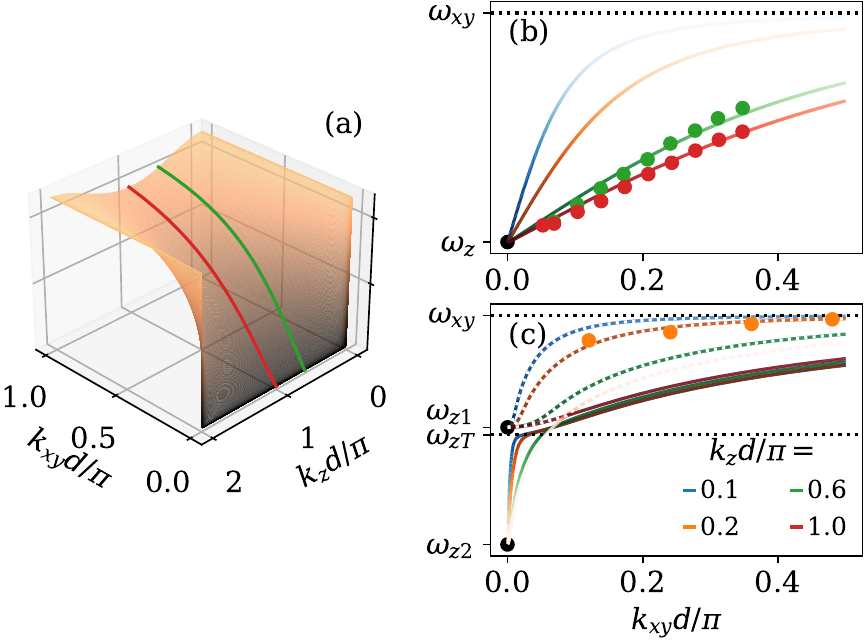}
		\caption{(a) Plasmon dispersion in a layered superconductor from Eq.\ \pref{eq:dis}. Red and green solid lines highlight cuts at fixed $k_z$. (b) RIXS measurements (dots) from Ref.\ \cite{nag_Phys.Rev.Lett.20} superimposed to plasmon branches plotted with a line intensity proportional to the THG polarization. In the bilayer case (c) the plasmon branches (shown in log scale) double, the upper one (dashed line) being the analogous of the single-layer mode. The lower branch, which rapidly approaches the $\omega_T$ scale seen in absorption (see text), is not visible in RIXS measurements \cite{bejas_Phys.Rev.B24}. The momentum scale and color codes are the same in panels (b) and (c).}
		\label{fig:1}
	\end{figure}

In layered superconductors, as e.g.\ cuprates,  the weak coupling between CuO$_2$ planes induces a marked anisotropy between in-plane $\omega_{xy}$ and out-of-plane $\omega_z$ long-wavelength plasma frequencies. While $\omega_{xy}$ lies at an energy scale where no appreciable change occurs across the SC transition, the soft $\omega_z$ plasmon becomes undamped by the opening of the SC gap below $T_c$, as evidenced by the emergence of a sharp plasma edge in the $c$-axis reflectivity at a temperature-dependent scale $\omega_z(T)$\cite{tamasaku_Phys.Rev.Lett.92,homes_Phys.Rev.Lett.93,kim_PhysicaC:Superconductivity95,basov_Phys.Rev.B94,vandermarel_CzechJPhys96,munzar_SolidStateCommunications99,kakeshita_Phys.Rev.Lett.01,dulic_Phys.Rev.Lett.01,hu_NatureMater14,dordevic_Phys.Rev.Lett.03}.  At finite momentum the two energy scales $\omega_{xy}$ and $\omega_z$ get mixed, and the energy-momentum dispersion of the layered three-dimensional (3D) plasmon, shown in Fig.\ \ref{fig:1}(a), acquires acoustic-like branches at fixed $k_z$, see Fig.\ \ref{fig:1}(b), due to the reduced screening between neighboring layers.  RIXS and EELS can probe such a dispersion both in the normal and SC states, see Fig.\ \ref{fig:1}(b) and (c), even though they cannot access the long-wavelength limit due to charge conservation\cite{hepting_Nature18,lin_npjQuantumMater.20,nag_Phys.Rev.Lett.20,bejas_Phys.Rev.B24}.

In cuprates pair tunnelling is the dominant inter-layer hopping mechanism below $T_c$, justifying a Josephson-like description for the out-of-plane phase modes\cite{savelev_Rep.Prog.Phys.10,savelev_NaturePhys06}. The excitation energy of SC Josephson plasmon  scales with the cosine of the SC phase gradient. Such an intrinsic anharmonicity turns in a primary source of non-linear optical response below $T_c$, thanks to the minimal coupling of the phase gradient to the gauge field\cite{savelev_Rep.Prog.Phys.10,savelev_NaturePhys06}. A typical hallmark of non-linear response is THz Third Harmonic Generation (THG). THG from Josephson plasmons\cite{rajasekaran_Science18,katsumi_Phys.Rev.B23,kaj_Phys.Rev.B23,zhang_NationalScienceReview23}, as well as additional effects connected to their non-linear response\cite{laplace_Adv.Phys.X16,fu_Phys.Rev.B22,liu_23}, have been experimentally proven in cuprates. In analogy with non-linear phononics\cite{juraschek_Phys.Rev.B18,vonhoegen_Nature18,forst_NaturePhys11,subedi_Phys.Rev.B14,li_Science19,basini_Phys.Rev.B24}, the main THG process can be understood as a sum-frequency process where two photons of the THz pulse centered at $\Omega$ excite simultaneously two plasma waves\cite{gabriele_NatCommun21} with opposite 3D momenta. As a consequence, while  linear reflectivity is only sensitive to the long-wavelength limits $\omega_{xy}$, $\omega_z$ of the plasmon dispersion, non-linear optics is sensitive to the convolution of plasmon excitations at {\em finite} momenta. This process retrieves the density of states of plasma excitations in a momentum region complementary to the one measured by RIXS, see Fig.\ \ref{fig:1}(b) and (c), providing in principle a mechanism to selectively tune the resonance energy for THz Third Harmonic Generation (THG) in a layered heterostructure. 

Here we demonstrate this principle by comparing single- and bi-layer structures. In single-layer systems, with a single plane per unit cell in the stacking direction, the light polarization projects out the full plasmon dispersion towards the low-energy $\omega_z$ value measured by linear optics. Since experiments are performed at a fixed pump frequency $\Omega$ as a function of temperature, the largest THG is expected at $\bar T$ where $\Omega=\omega_z(\bar T)$. Such a prediction, qualitatively similar to what obtained\cite{gabriele_NatCommun21} by neglecting the plasmon dispersion, is consistent with experiments \cite{rajasekaran_Science18, kaj_Phys.Rev.B23}. 

In bilayer superconductors  the modulation of intra-bilayer $J_{z1}$ and inter-bilayer $J_{z2}$ Josephson couplings along the stacking direction, with $J_{z1}\gg J_{z2}$, leads to a doubling of the Josephson plasmon branches, as originally explained in Ref.\ \cite{vandermarel_CzechJPhys96}, giving in the long-wavelength limit two scales $\omega_{z1}\gg \omega_{z2}$. The lower one $\omega_{z2}\propto \sqrt{J_{z2}}$ controls the lower and sharper reflectivity edge at zero momentum\cite{homes_Phys.Rev.Lett.93,kim_PhysicaC:Superconductivity95,basov_Phys.Rev.B94,vandermarel_CzechJPhys96,kakeshita_Phys.Rev.Lett.01,dulic_Phys.Rev.Lett.01,hu_NatureMater14}, while the upper Josephson branch has been recently measured by RIXS\cite{bejas_Phys.Rev.B24}, that turns out to be insensitive to the lower plasmon branch, see Fig.\ \ref{fig:1}(c). By computing the THG for the bilayer case we show  that the dominant contributions of plasma modes to the THG arise from intra-bilayer phase fluctuations, scaling with the larger Josephson stiffness. As a consequence, in the bilayer system the momentum integration moves the THG spectral weight towards a higher energy scale $\omega_{zT}\approx \omega_{z1}$, explaining  the lack of any resonance at the temperature $\bar T$ where $\Omega=\omega_{z2}(\bar T)$ in the measurements of Ref.\ \cite{katsumi_Phys.Rev.B23}.

\begin{figure}
		\centering
		\includegraphics[width=\columnwidth]{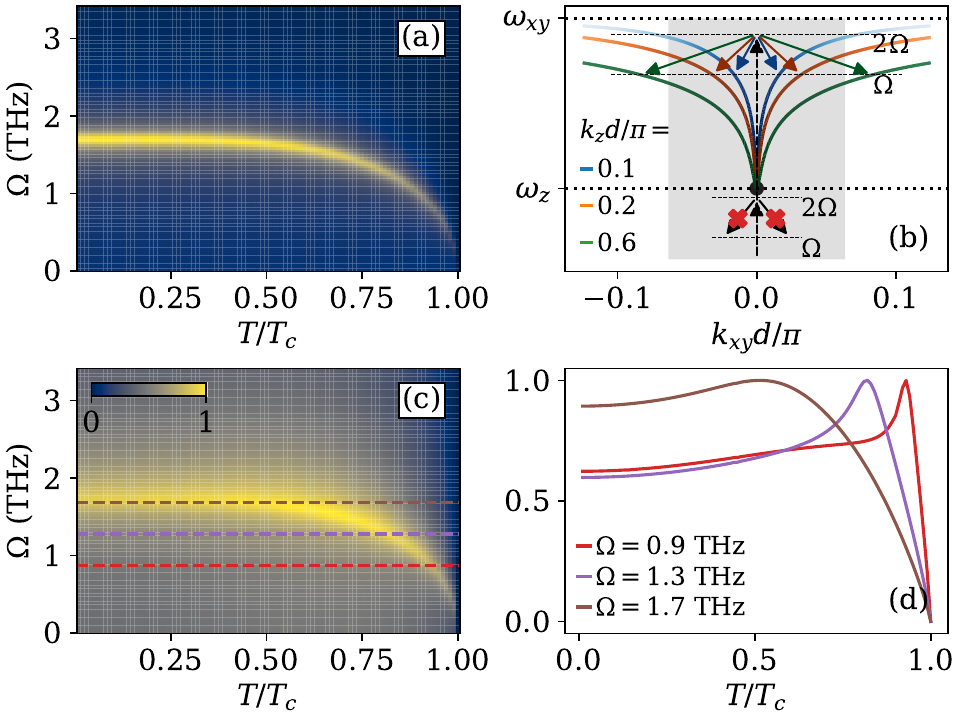}
		\caption{Comparison between the nonlinear kernel $\lvert K(2\Omega,T)\rvert$ of Eq.\ \pref{eq:ker} for (a) 
		non-dispersive plasmon, corresponding to replacing $\omega_L(\bk)$ with $\omega_z$, and (c) the dispersive plasmon. (b) Sketch of the THG process. The $\omega_L(\bk)$ dispersion from Eq.\ \pref{eq:dis} is plotted in log scale, with a line intensity weighted with the spectral weight of the two-plasmon process from Eq.\ \pref{eq:ker}.  Only for $\Omega>\omega_z$ two light pulses at $\Omega$ can decay in two plasmons with opposite momenta. At larger energies the number of available states increases, but these correspond to a region with smaller spectral weight (see text and \cite{suppl}). The shaded area corresponds to the region $k_{xy}<1/\xi$ of allowed in-plane momenta. (d) $\lvert K(2\Omega)\rvert$ from panel (c) as a function of temperature for three selected pump frequencies (dashed lines in panel (c)), where each curve is normalized to its maximum.}
		\label{fig:2}
	\end{figure}

{\em Single layer case}.	 
	The contribution of plasma waves to the non-linear optical response can be qualitatively understood within a simplified anisotropic Josephson-like model for the SC phase. Labelling with $\theta_n$ the SC phase in the $n$-th layer one can write an effective classical Hamiltonian:
	\begin{equation}
	\lb{eq:hs}
	H=\frac{1}{8}\int d^2r\,d\sum_n\left[D_{xy}(\nabla_{xy}\theta)^2-8J_z\cos(\theta_{n+1}-\theta_n)\right],
	\end{equation}
	where the in-plane phase difference has been already promoted to a continuum phase gradient, $D_{xy}$ is the in-plane stiffness (equivalent to $\hbar^2 n_s/m$ in the continuum, isotropic limit) and $J_z$ is an energy density defining the superfluid stiffness as $D_z=4J_zd^2$ along the out-of-plane $z$ direction, $d$ being the interlayer distance. Henceforth, unless explicitly displayed, we set $\hbar=k_B=1$. The classical model \pref{eq:hs} is promoted to a quantum equivalent by including dynamical effects related to the phase gradient in time, scaling as $\kappa_0(\partial_\tau \theta_n)^2$, with $\kappa_0$ the charge compressibility. Once that Coulomb effects are included, $\kappa_0$ is replaced by the RPA charge  compressibility\cite{benfatto_Phys.Rev.B04,sun_Phys.Rev.Res.20}. By retaining only the quadratic term in the expansion of the cosine in Eq.\ \pref{eq:hs} we obtain the following Gaussian quantum action\cite{suppl,paramekanti_Phys.Rev.B00,depalo_Phys.Rev.B99}:
	\begin{equation}
	\lb{eq:sg}
	S_G=\frac{1}{32\pi e^2}\sum_{i\omega_m,\mathbf{k}}\lvert\mathbf{k}\rvert^2\left(-(i\omega_m)^2+\omega_L^2(\mathbf{k})\right)\lvert\theta(i\omega_m,\mathbf{k})\rvert^2,
	\end{equation}
	where $i\omega_m = 2\pi m T$ are bosonic Matsubara frequencies, and $\omega_L(\mathbf{k})$ includes the full momentum dependence of the layered Josephson plasma mode (JPM):
	\begin{equation}
	\lb{eq:dis}
	\omega_L^2(\mathbf{k})=\omega_{xy}^2\frac{k_{xy}^2}{\lvert \mathbf{k}\rvert^2}+\omega_{z}^2\frac{q_{z}^2}{\lvert \mathbf{k}\rvert^2},
	\end{equation}
	where $\omega_{xy,z}^2=4\pi e^2D_{xy,z}$, $q_z=(2/d)\sin{(k_zd/2)}$  and $|\bk|^2=k_{xy}^2+q_z^2$. An external gauge field $A_z$ polarized along the $z$ direction enters the model \pref{eq:hs} via the minimal coupling substitution $\theta_{n+1}-\theta_n \rightarrow \theta_{n+1}-\theta_n -2edA_z/c$. As detailed in the Supplementary\cite{suppl}, the third-order current $J^{NL}\sim KA_z^3$ can be obtained by expanding the cosine Josephson interaction in Eq.\ \pref{eq:hs} beyond second order in the gauge-invariant phase gradient along $z$, and integrating out the phase degrees of freedom\cite{gabriele_NatCommun21}. For a monochromatic incident field $A_z=A_0\cos(\Omega t)$ the resulting THG intensity is proportional\cite{cea_Phys.Rev.B16,udina_Phys.Rev.B19,seibold_Phys.Rev.B21} to $I_{THG}\propto |K(2\Omega)|^2$. The resonant part of the non-linear kernel $K$ accounts for the density-of-states of the process involving the generation of two JPMs with opposite momenta, see Fig.\ \ref{fig:2}(b), and it is given explicitly by:
	\begin{equation}
	\lb{eq:ker}
	K(\Omega)\propto J_z^2\frac{1}{V}\sum_{\mathbf{k}}\frac{q_z^4}{\lvert\mathbf{k}\rvert^4}\frac{\coth{(\frac{\beta\omega_L(\mathbf{k})}{2})}}{\omega_L(\mathbf{k})(4\omega^2_L(\mathbf{k})-(\Omega+i\delta)^2)},
	\end{equation}
	where $i\omega_m\rightarrow \Omega+i\delta$. Within the Josephson model \pref{eq:hs} the coupling $J_z$ controls both the stiffness and the anharmonic interaction terms. Thus we will adopt the same experimental temperature-dependent stiffness $D_z(T)$, $D_{xy}(T)$  both in the plasmon dispersion \pref{eq:dis} and in the effective anharmonic couplings, see\cite{suppl,shibauchi_Phys.Rev.Lett.94}. In addition, we show in the Supplementary\cite{suppl} that the effects of the mixing\cite{gabriele_Phys.Rev.Res.22,sellati_Phys.Rev.B23,gabriele_Phys.Rev.B24} between longitudinal (plasmon) and transverse (polariton) branches below a typical scale $k_c\sim \sqrt{\omega_{xy}^2-\omega_z^2}/c$ is quantitatively irrelevant for the THG,  since the in-plane integration extends up to a much larger scale $k_{xy}\sim 1/\xi$, where $\xi\sim 5d$ is the in-plane coherence length\cite{ando_Phys.Rev.Lett.02,suzuki_Phys.Rev.B91}.


	In Fig.\ \ref{fig:2} we show the nonlinear kernel $K$ as a function of temperature and frequency, by setting $\omega_{xy}(T=0)=250$ THz and $\omega_{z}(T=0)=1.7$ THz, as appropriate for optimally-doped LSCO samples ($T_c=38$ K)\cite{kaj_Phys.Rev.B23}. 
We compare the case where the full plasmon dispersion \pref{eq:dis} is included (panel c) with the case, considered previously\cite{gabriele_NatCommun21}, where $\omega_L(\bk)$ is assumed to be non dispersive (panel a). The latter case is equivalent to setting $\omega_L(\bk)=\omega_z$ in Eq.\ \pref{eq:ker}, which results in a resonance at $\Omega=\omega_z$ of the kernel $K(2\Omega)$. As one can see, even though the dispersive case leads to a smearing of such a sharp resonance, the non-linear kernel shows nonetheless a marked maximum around $\omega_z$, see panel (c). The reasons are the following. As detailed in \cite{suppl}, the spectral weight of the two-plasmon process is zero when $\Omega < \omega_z$, with a discontinuity at the edge, which consequently causes a singularity in $\lvert K(2\Omega=2\omega_z)\rvert$. Secondly, the overall factor $q_z^4/|\bk|^4$ in Eq.\ \pref{eq:ker}, which accounts for the polarization of the light in the $z$ direction, projects out  the full dispersion  \pref{eq:dis}  on the phase space where $q_z/|\bk|$ is large, that corresponds to an energy around $\omega_z$. In addition, the resonance at $\omega_L$ in Eq.\ \pref{eq:ker} scales overall as $1/\omega_L^2$, highlighting further low-momentum processes, i.e. those less relevant for RIXS\cite{hepting_Nature18,lin_npjQuantumMater.20,nag_Phys.Rev.Lett.20}, see Fig.\ \ref{fig:1}(b). The combined action of these effects is visualized in Fig. \ref{fig:2}(b), where the intensity of each plasmon line is proportional to its contribution to the non-linear kernel \pref{eq:ker}, and further discussed in \cite{suppl}. The overall enhancement of the kernel at energies around $\omega_z$ leads to a non-monotonic $T$ dependence of the THG measured at a fixed pump frequency $\Omega<\omega_{z}(T=0)$, see Fig.\ \ref{fig:2}(d), maximum at $\bar T$ where $\Omega=\omega_z(\bar T)$, in qualitative agreement with the result obtained for a non-dispersive plasmon\cite{gabriele_NatCommun21}. Notice that even if the two-plasmon process is only possible above $2\omega_z$, the real part of the optical kernel \pref{eq:ker} is finite below it, giving THG even when the pump frequency is below $\omega_z(T)$.

	\begin{figure*}
		\centering
		\includegraphics[width=2\columnwidth]{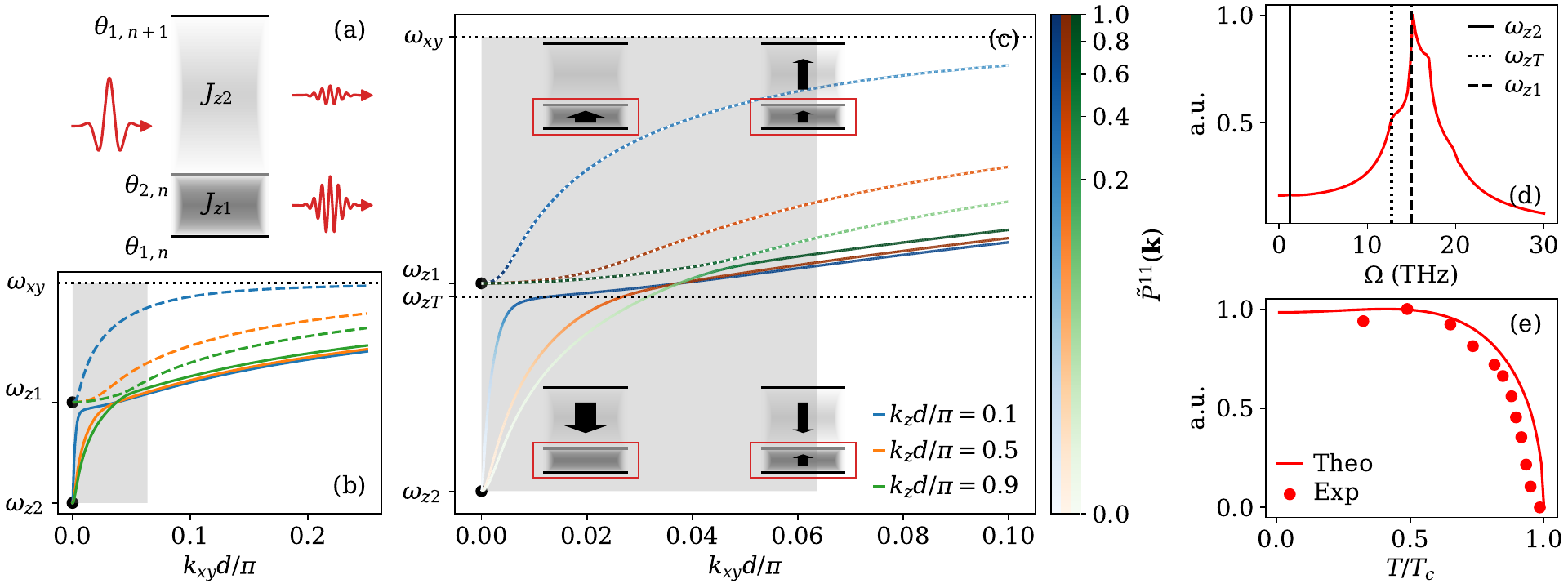}
		\caption{(a) Sketch of the $c$-axis non-linear response for the YBCO bilayer system. Since $J_{z1}\gg J_{z2}$ the THG signal is mainly determined by the intra-bilayer phase fluctuations. (b) 	
		Plasmon branches $\omega_{1}(\bk)$ (dotted lines) and $\omega_{2}(\bk)$ (solid lines) at selected $k_z$ for a wide range of $k_{xy}$ value. (c) Zoom-in of panel (c) in the region of integration relevant for the THG, as denoted by the grey shaded area. Here the line intensity is rescaled according to the relative weight of intra-bilayer phase fluctuations $\tilde{\mathcal{P}}_{\sigma}^{11}$ of each mode, shown as a color-bar  on the right of the panel. A sketch of the corresponding polarizations of the modes\cite{sellati_Phys.Rev.B23}  in each momenta region is  also reported. (d) Frequency dependence of the non-linear kernel $\lvert K(2\Omega)\rvert$ at $T=0$. Vertical bars denote the position of $\omega_{z1}$, $\omega_{zT}$ and $\omega_{z2}$. (e) Temperature dependence of the non-linear kernel $|K(2\Omega)|$  (solid line) for fixed pump frequency $\Omega=0.5$ THz. The dots represents the experimental data  from Ref.\ \cite{katsumi_Phys.Rev.B23}.}
		\label{fig:3}
	\end{figure*}
	
{\em Bilayer case.}	
	In the bilayer case one has two planes denoted as $\lambda=1,2$ per unit cell, so that one introduces two phase variables $\theta_{\lambda,n}$. The out-of-plane hamiltonian density is
	\begin{equation}
	\lb{eq:oop}
	\mathcal{H}^{\perp}_n=-J_{z1}\cos(\theta_{1,n}-\theta_{2,n})-J_{z2}\cos(\theta_{1,n+1}-\theta_{2,n}),
	\end{equation}
	in close analogy with Eq.\ \pref{eq:hs}, while the in-plane part has the same form. Here we introduced two different Josephson couplings $J_{z1}$ and $J_{z2}$, with $J_{z1}\gg J_{z2}$, corresponding to a phase difference  between the two nearest planes within the same unit cell (at distance $d_1=3.2\,\angstrom$ for YBCO) or in neighboring cells (at distance $d_2=8.2\,\angstrom$ for YBCO), $d=d_1+d_2$ being the periodicity in the $z$ direction\cite{sellati_Phys.Rev.B23,alpeggiani_Phys.Rev.B13,vandermarel_CzechJPhys96}, see Fig.\ \ref{fig:3}(a). The corresponding stiffnesses $D_{z\lambda}=4J_{z\lambda}d_{\lambda}^2$ are also modulated. By introducing the spinor $\bar{\theta}=(\theta_{1},\theta_{2})$ the generalization of the Gaussian quantum action \pref{eq:sg} to the bilayer case reads\cite{suppl}
	\begin{equation}
	\label{eq:sgbi}
	S_G=\frac{1}{64\pi e^2}\sum_{i\omega_m,\mathbf{k}}\bar{\theta}(-k)\mathcal{K}^2\left(-(i\omega_m)^2\mathbb{1}+\Omega^2_L(\mathbf{k})\right)\bar{\theta}(k).
	\end{equation}
	Here we introduced the $2\times2$ matrix $\mathcal{K}^2=k_{xy}^2\mathbb{1}+\mathcal{K}_z^{\dagger}\mathcal{K}_z$ which generalizes the quadratic term in momentum, where 
	\begin{equation}
	\mathcal{K}_z=\frac{\sqrt{2}}{d}
	\begin{pmatrix}
		-\frac{e^{-ik_zd_1/2}}{\sqrt{d_1/d}}&\frac{e^{ik_zd_1/2}}{\sqrt{d_1/d}}\\ \frac{e^{ik_zd_2/2}}{\sqrt{d_2/d}}&-\frac{e^{-ik_zd_2/2}}{\sqrt{d_2/d}}
	\end{pmatrix},
	\end{equation}
	keeps track of the discretization of the phase variables along the $z$ direction. Analogously $\Omega^2_L(\mathbf{k})$ generalizes the plasma dispersion \pref{eq:dis}, with $\Omega^2_L(\mathbf{k})=(\mathcal{K}^{2})^{-1}(\omega_{xy}^2k_{xy}^2\mathbb{1}+\mathcal{K}_z^{\dagger}\Omega_z^2\mathcal{K}_z)$, where  $(\Omega_z^2)_{\lambda\mu}=\delta_{\lambda\mu}\omega_{z\lambda}^2$ and $\omega_{z\lambda}^2=4\pi e^2 D_{z\lambda}$. From the structure of Eq.\ \pref{eq:sgbi} one readily sees that the doubling of the layers per unit cell leads to a doubling of the JPMs, defined as the square-root of the eigenvalues of the matrix $\Omega^2_L(\mathbf{k})$, and shown in Fig.\ \ref{fig:3}(b). The upper plasmon $\omega_1(\bk)$, dispersing between $\omega_{z1}$ and $\omega_{xy}$, is the equivalent of the single-layer plasmon $\omega_L(\bk)$, while the additional low-energy out-of-plane plasmon $\omega_2(\bk)$, dispersing away from $\omega_{z2}$, is characteristic of the bilayer case. For increasing in-plane momentum $\omega_2(\bk)$ evolves rapidly towards an energy scale $\omega_{zT}=\sqrt{(\omega_{z1}^2d_2+\omega_{z2}^2d_1)/d}\approx\omega_{z1}$ for $J_{z1}\gg J_{z2}$, usually named \emph{transverse} plasmon in the literature because it identifies the scale of a resonance peak in the $c$-axis optical conductivity \cite{vandermarel_CzechJPhys96,munzar_SolidStateCommunications99,hu_NatureMater14}. For even larger values of $k_{xy}$ $\omega_2(\bk)$ grows slowly towards the in-plane frequency $\omega_{xy}$.
		
The computation of the non-linear kernel follows the same steps as before, with both plasmon branches contributing to the 
out-of-plane response, so that  
$K(\Omega)=K_{+}(\Omega)+K_{-}(\Omega)$, where 
	\begin{equation}
	\label{eq:kerbi}
	K_{\pm}(\Omega)=\sum_{\sigma\sigma^{\prime},\mathbf{k}}W_{\sigma\sigma^{\prime}}
	\frac{\omega_{\sigma}\pm\omega_{\sigma^{\prime}}}{4\omega_{\sigma}\omega_{\sigma^{\prime}}}\frac{\coth{(\frac{\beta\omega_{\sigma}}{2})}\pm\coth{(\frac{\beta\omega_{\sigma^{\prime}}}{2})}}{(\omega_{\sigma}\pm\omega_{\sigma^{\prime}})^2-(\Omega+i\delta)^2}
	\end{equation}
	combines two JPMs from the same ($\sigma=\sigma^{\prime}$) or from different ($\sigma\neq\sigma^{\prime}$) plasmon branches, with $\sigma,\sigma^{\prime}=1,2$, weighted by the $\mathbf{k}$-dependent $W_{\sigma\sigma^{\prime}}$ prefactors, which project out  the contribution of each mode to selected region of momenta. 
	A full numerical computation of the kernel \pref{eq:kerbi} at $T=0$ for parameter values  appropriate for YBCO is shown in Fig.\ \ref{fig:3}(d). We set $T_c=61$ K, $\omega_{z1}(T=0)=15$ THz and $\omega_{z2}(T=0)=1.2$ THz\cite{katsumi_Phys.Rev.B23,hu_NatureMater14}, that gives a peak in the transverse conductivity at $\omega_{zT}(T=0)=11$ THz, in agreement with the experiments\cite{hu_NatureMater14}. 
	As one can see, $\lvert K(2\Omega)\rvert$ shows a marked resonance near the value $\Omega\approx \omega_{z1}\sim15$ THz, without any signature at the lower plasmon $\omega_{z2}\sim1.2$ THz, that is the energy scale defining  instead the sharp plasma edge in the $c$-axis reflectivity below $T_c$\cite{homes_Phys.Rev.Lett.93,basov_Phys.Rev.B94,vandermarel_CzechJPhys96,hu_NatureMater14}. 
	
	A better insight into this result can be obtained by introducing a new sets of phase variables, corresponding to the intra-bilayer $\tilde{\theta}_{1}\sim\theta_{2,n}-\theta_{1,n}$ and inter-bilayer $\tilde{\theta}_{2}\sim\theta_{1,n+1}-\theta_{2,n}$  phase gradient in the $n$-th unit cell, respectively. As manifest from \pref{eq:oop}, the larger scale $J_{z1}$ couples the gauge field $A_z$ to $\tilde{\theta}_1$, while $J_{z2}$ controls its coupling to $\tilde{\theta}_2$. By changing basis from $(\theta_{1},\theta_{2})$ to $(\tilde{\theta}_{1},\tilde{\theta}_{2})$  one can write\cite{suppl} the projectors as $W_{\sigma\sigma^{\prime}}\propto\Tr{(\tilde{\mathcal{P}}_{\sigma}\mathcal{C}\tilde{\mathcal{P}}_{\sigma^{\prime}}\mathcal{C})}$, where the diagonal matrix $(\mathcal{C})_{\lambda\mu}=D_{z\lambda}d_\lambda^2\delta_{\lambda\mu}$ comes from the coupling with the gauge field ($\sim D_{z\lambda}A_z^2{d_\lambda}^2$) and the matrix $\tilde{\mathcal{P}}_{\sigma}$ weights each plasmon branch according to its projection onto inter-bilayer ($11$ element) and intra-bilayer ($22$ element) phase fluctuations. Since $D_{z1}\gg D_{z2}$, one finds  that plasmon branches contribute to THG in the momentum region where they have a large intra-bilayer component, $\tilde{\mathcal{P}}_{\sigma}^{11}(\bk)$. This effect is highlighted in Fig.\ \ref{fig:3}(c), where the line intensity for each plasmon dispersion, plotted at fixed $k_z$, scales with its $\tilde{P}_{\sigma}^{11}(\bk)$ intra-bilayer weight. The upper  plasmon branch $\omega_1(\mathbf{k})$ is purely intra-bilayer for $k_{xy}=0$, and gradually acquires an inter-bilayer component as $k_{xy}$ increases. Thus, it contributes to THG in the low-momenta region where $\omega_1(\bk)\approx \omega_{z1}$, as already observed for the $\omega_L(\bk)$ in the single-layer case. Conversely, the lower plasmon branch $\omega_2(\mathbf{k})$ has exactly zero intra-bilayer projection at $k_{xy}=0$, where it is associated with purely inter-bilayer phase fluctuations\cite{sellati_Phys.Rev.B23,suppl}, while it acquires intra-bilayer character as it approaches the $\omega_{zT}$ energy scale. The suppression of spectral weight at low energy comes from the deep entanglement between the momentum dependence of the plasmon energies and of the polarization weights, which would not be captured if the system were described neglecting the dispersion of the plasmon branches. These two ingredients explain the lack on any signature at $\omega_{z2}$  in the numerical result of Fig.\ \ref{fig:3}(d) at $T=0$, and the absence of a THG  enhancement at $\Omega=\omega_{z2}(T)$ in the THG measurements Ref.\ \cite{katsumi_Phys.Rev.B23},  performed with fixed $\Omega\sim 0.5$ THz. At the same time the condition $\Omega=\omega_{z1}(\bar T)$ is achieved at a temperature $\bar T$ so close to $T_c$ that the resonance itself is washed out by the thermal depletion of the stiffness. Thus the computed THG, given by the solid line in Fig.\ \ref{fig:3}(e), increases monotonically in temperature, in remarkable agreement with the experimental THG of Ref.\ \cite{katsumi_Phys.Rev.B23}, extracted by the measured reflected electric field $E_r$  as\cite{fre}:
	\begin{equation}
		\lb{eq:er}
		E_{r}\propto K(2\Omega)\Bigl(\frac{2}{n(\Omega)+1}\Bigr)^3\frac{1}{n(3\Omega)+n(\Omega)}\frac{1}{n(3\Omega)+1}E_0^3,
	\end{equation}
	where $n(\Omega)$ $(n(3\Omega))$ is the experimentally-measured index of refraction at $\Omega$ $(3\Omega)$. Eq.\ \pref{eq:er} is a modified Fresnel result which includes also the role played by the non-linear current\cite{fre}.

{\em Conclusions}	
	The present results for the two-plasmon contribution to the non-linear response reconcile experiments in single-layer and bilayer cuprates. The combined effect of the light polarization and the plasmon dispersion explains a resonant THG for pumping frequency matching the plasma edge in single-layer materials, while it accounts for a shift of the resonance frequency towards the larger out-of-plane plasma mode for bilayer structures. It turns out that in those systems THG is sensitive exactly to the part of the plasmon branches that are hidden to RIXS, as shown in Fig\ \ref{fig:1}(c), where recent RIXS measurements from Ref.\ \cite{bejas_Phys.Rev.B24} have been reported. Even though such a dichotomy is not yet understood theoretically\cite{bejas_Phys.Rev.B24}, we expect that it should come from different selection rules at play for the plasma-mode contribution to the current (probed by THG) or to the density (probed by RIXS) response.

	On a more general perspective, the present results can serve as a guide to engineer artificial heterostructures for the generation of THz pulses via THG up-conversion, using the plasmon dispersion to tune the enhancement of the non-linear response for pumping THz field polarized along the layer stacking direction in cuprate superconductors. In the last few years giant progresses have been made in the realization of stable superconducting films of Bi-based cuprates, down to the dimensions of few unit cells\cite{yu_Nature19,zhao_Phys.Rev.Lett.19,poccia_Phys.Rev.Mater.20}. It is now possible to realize efficient and tunable Josephson junctions by twisting neighbouring films, thanks to the angle-dependence of the pair tunnelling in a $d$-wave superconductor\cite{zhao_Science23, lee_Adv.Mater.23,martini_MaterialsToday23}.  Even though the present technology is still limited to single junctions, in the near future the realisation of Junction arrays could lead to controllable artificial bilayer lattices, with a tunable resonance frequency. At the same time, the full quantum description of the non-linear Josephson response is the starting point for the understanding of the multidimensional spectrum obtained in two-dimensional THz protocols\cite{liu_23}, and to asses its potential for disentangling the intrinsic thermal effects from inhomogeneity broadening in unconventional cuprates.

	{\em Acknowledgements.} We acknowledge K.\ Katsumi  and N.\ Poccia for useful discussions. We thank K.\ Katsumi for providing us with the experimental data from Ref.\ \cite{katsumi_Phys.Rev.B23}. We acknowledge financial support by  European Community by ERC-SYN MORE-TEM G.A. No 951215 and by Sapienza University via Ateneo 2021 (RM12117A4A7FD11B), Ateneo 2022 (RP1221816662A977) and Ateneo 2023 (RM123188E357C540).

\bibliography{thg_plasmons_bib.bib}

\end{document}